\documentclass[prd,11pt,nofootinbib]{revtex4-1}
\usepackage[mathcal]{euscript}
\usepackage{latexsym}
\usepackage{amsmath}
\usepackage{hyperref}
\usepackage{graphicx}   
\usepackage{verbatim}   
\usepackage{color}      
\usepackage{subfigure}  
\usepackage{hyperref}   
\usepackage{bm}
\usepackage{graphicx}
\usepackage{amssymb}
\usepackage{bbm}
\usepackage{float}
\usepackage{slashed}
\usepackage{amsfonts}
\usepackage{euscript}
\usepackage{mathrsfs} 
\usepackage{bbding}
\usepackage{pifont}
\usepackage{cancel}

\makeindex

\begin{document}

\title{Where does the physics of extreme gravitational collapse reside?}

\author{Carlos Barcel\'o}
\email{carlos@iaa.es}
\affiliation{Instituto de Astrof\'{\i}sica de Andaluc\'{\i}a (IAA-CSIC), Glorieta de la Astronom\'{\i}a, 18008 Granada, Spain}
\author{Ra\'ul Carballo-Rubio}
\email{raulc@iaa.es}
\affiliation{Instituto de Astrof\'{\i}sica de Andaluc\'{\i}a (IAA-CSIC), Glorieta de la Astronom\'{\i}a, 18008 Granada, Spain}
\affiliation{Departamento de Geometr\'{\i}a y Topolog\'{\i}a, Facultad de Ciencias, Universidad de Granada, Campus Fuentenueva, 18071 Granada, Spain}
\author{Luis J. Garay}
\email{luisj.garay@ucm.es}
\affiliation{Departamento de F\'{\i}sica Te\'orica II, Universidad Complutense de Madrid, 28040 Madrid, Spain}
\affiliation{Instituto de Estructura de la Materia (IEM-CSIC), Serrano 121, 28006 Madrid, Spain}

\begin{abstract}{The gravitational collapse of massive stars serves to manifest the most severe deviations of general relativity with respect to Newtonian gravity: the formation of horizons and spacetime singularities. Both features have proven to be catalysts of deep physical developments, especially when combined with the principles of quantum mechanics. Nonetheless, it is seldom remarked that it is hardly possible to combine all these developments into a unified theoretical model, while maintaining reasonable prospects for the independent experimental corroboration of its different parts. In this paper we review the current theoretical understanding of the physics of gravitational collapse in order to highlight this tension, stating the position that the standard view on evaporating black holes stands for. This serves as the motivation for the discussion of a recent proposal that offers the opposite perspective, represented by a set of geometries that regularize the classical singular behavior and present modifications of the near-horizon Schwarzschild geometry as the result of the propagation of non-perturbative ultraviolet effects originated in regions of high curvature. We present an extensive exploration of the necessary steps on the explicit construction of these geometries, and discuss how this proposal could change our present understanding of astrophysical black holes and even offer the possibility of detecting genuine ultraviolet effects on future gravitational wave experiments.
}
\end{abstract}


\maketitle
\flushbottom
\tableofcontents

\section{Introduction \label{sec:1}}

Black holes are currently accepted as members of the bestiary of astronomical objects. Their success is explained on the one hand by the simplicity and elegance of their mathematical properties, and on the other hand by the (indirect) observation of the existence of very dense and intrinsically very dark distributions of matter in our universe. For instance, observations of the center of our own galaxy point to the existence of an object with such characteristics that the only known theoretical structure in general relativity (GR) that can be identified with it is a supermassive black hole \cite{Schoedel2002,Schoedel2003,Ghez2005}. However, it is noteworthy that we lack direct evidence of the most characteristic property of a black hole, namely its horizon (this may actually be impossible \cite{Abramowicz2002}; for a short discussion of this issue see Sec.~\ref{sec:6a}), or even of the very dynamical formation of black holes. The latter deficiency could eventually be alleviated in the future with the advent of gravitational wave astronomy \cite{Fryer2011}.

Moreover, black holes contain odd features that suggest that their theoretical portrait is far from completely understood. The most prominent of these features is the unavoidable presence of a singularity, a region in spacetime in which the known laws of physics break down \cite{Hawking1976,Hawking1973}. It is expected that the successful combination of quantum mechanics and gravity will lead to the regularization of these singularities. In the absence of a full theory of quantum gravity or, in a wider sense, of an ultraviolet completion of GR, there has nonetheless emerged, especially after the work of Hawking \cite{Hawking1975,Hawking1976}, a consensus picture about the kind of ultraviolet modifications of the classical behavior one should expect. Naturally, the overall picture is far from being completely self-consistent (see for instance the information loss problem \cite{Hawking2005} or the recent firewall controversy \cite{Almheiri2013}), and surprises may arise as our knowledge about the high-energy properties of the gravitational interaction improve.

This is the vision we align with in this paper. Guided by the discomfort with a ubiquitous property of the models that are currently in the market, namely the absurdly large lifetime of evaporating black holes, we elaborate on a completely different (and in some sense to be detailed, opposite) view on the possible ultraviolet effects on gravitational collapse processes, that was first discussed in \cite{Barcelo2011,Barcelo2014e,Barcelo2015}. Quite concisely, our geometric construction represents the transition from a black-hole geometry to a white-hole geometry in a short characteristic time scale. It is not our aim to provide here a first-principle justification of the overall proposal, though we shall include some thoughts on promising developments to consider in this direction. Our approach is heuristic, trying to elucidate what kind of effects have the potential of being eventually measurable, even if the evaluation of this potential could be at the moment rather theoretical. In a genuine bottom-up approach, we shall focus in the self-consistency of the picture and the implications that derive from it, for both the high-energy theoretical aspects and the low-energy observational features.

This paper complements our previous discussions, presenting a more detailed account of some of the motivations behind this model and an in-depth exploration of its main geometric and physical features. The body of the paper is divided in five main sections. In Secs. \ref{sec:2} and \ref{sec:3} we review, respectively, some of the basic features of gravitational collapse of massive stars and properties of black holes in GR, and the quantum-mechanical modifications of some of these properties that are nowadays contemplated from a theoretical perspective. In Sec. \ref{sec:4} we look at the different possible ultraviolet modifications from a phenomenological perspective. In Sec. \ref{sec:5} we discuss our proposal in detail, presenting a step-by-step construction of a family of geometries describing the process we want to describe. In Sec. \ref{sec:6} we go through the possible observational implications of the proposal, that may be of relevance if these processes are indeed realized in nature. We finish with a brief conclusions section.

\section{Black holes and spacetime singularities: the standard GR picture \label{sec:2}}

\subsection{Event horizons \label{sec:2a}}

We shall concentrate on the simplest models of gravitational collapse, which are spherically symmetric. Eventually we will consider the effect of introducing small, non-spherical perturbations in this simple picture. As is well known, in the spherically symmetric case, and for a star with mass $M$ whose radius is larger than its Schwarzschild radius,
\begin{equation}
r_{\rm s}=\frac{2GM}{c^2},
\end{equation}
the external metric is unique and is given precisely by the Schwarzschild solution. The Schwarzschild metric has to be glued with the internal metric of the matter distribution along the surface of the star. A stable stellar structure has to be in hydrostatic equilibrium described by means of the Tolman-Oppenheimer-Volkoff equation \cite{Oppenheimer1939}. On the basis of this equation and known equations of state for the matter inside the star, one concludes that there exists an upper bound to the mass of stable configurations \cite{Oppenheimer1939,Rhoades1974,Bombaci1996}. Accordingly, any star in equilibrium will become unstable if accreting more than a certain amount of matter. 

In those extreme situations, the strength of the gravitational interaction surpasses that of any other possible known force in the system, provoking an indefinite contraction of the stellar structure. The resulting trajectory of the surface of the collapsing star can be described by means of a function $R(\tau)$, where $\tau$ is the proper time of an observer attached to the surface. If the regime in which the strength of the gravitational interaction surpasses that of any other relevant force in the system is reached, this surface will inevitably cross the Schwarzschild radius in finite time. In terms of equations, for a given initial stellar radius at $\tau=\tau_{\rm i}$ there always exists a finite proper time interval $\Delta\tau$ so that
\begin{equation}
R(\tau_{\rm i}+\Delta\tau)=r_{\rm s}.
\end{equation}
A prime example of this behavior is the Oppenheimer-Snyder model of gravitational collapse \cite{Oppenheimer1939b}, in which the perfect fluid representing the matter content is pressureless, so that gravity is strictly the only force present in the model. The Oppenheimer-Snyder model is thus considered to provide a reliable description of the late stages of the gravitational collapse process for massive enough distributions of matter with small asphericities.

Upon crossing, a trapping horizon is formed in the position that corresponds to the Schwarzschild radius. While we leave to the next section the rigorous definition of such a notion, for the moment it is enough to have in mind the intuitive picture that this trapping horizon \emph{locally} forbids that signals originated at the surface of the star reach external observers. This can be seen explicitly if we take the black-hole patch of the Kruskal manifold and use suitable coordinates, for instance Painlev\'e-Gullstrand coordinates (see \cite{Martel2001} and references therein). It is important to keep in mind that this observation by itself does not imply that lumps of matter will never be able to cross outwards the radial position at which the trapping horizon was first formed. The overall dynamical evolution of spacetime around the outgoing distribution of matter should be taken into account in order to consistently conclude so. As we will see, restrictions to the geometry of spacetime in the form of energy conditions are useful to tackle this issue. We leave this discussion for the following section, assuming for the moment the standard picture that horizons are inviolable in classical GR.

The resulting object, a so-called black hole, would then possess as defining characteristic an event horizon. They would be absolutely inert objects, the ultimate end point of gravitational collapse, the dead state of stellar physics. No matter content can ever escape from the inside, as the region behind the Schwarzschild radius is causally disconnected from the external observers. Black holes are characterized by just three numbers, mass $M$, electric charge $Q$ and angular momentum $J$, for generic initial conditions (even if not spherically symmetric) for the matter that went to form the black hole; see \cite{Chrusciel2012} for a thorough technical discussion. This simplicity is an appealing feature for a large number of theorists. In our idealized setting we are demanding spherical symmetry ($J=0$) and electrical neutrality ($Q=0$), leaving us with just one relevant parameter.

\subsection{Singularities \label{sec:2b}}

According to classical GR, the fate of matter after crossing the trapping horizon is ominous. While the horizon is characterized by a pronounced deformation of the light cones so that (locally) the light that is emitted at the horizon cannot escape, light cones just behind the horizon are deformed even more drastically. The resulting deformation ultimately implies that every light-like trajectory starting inside the horizon must (locally) go inwards, i.e., with the value of the radial coordinate decreasing, until ``hitting'' the singularity.

Let us be more precise on these notions, which not only will permit us to make sharp statements, but also understand the generality of this picture. In principle one could think (and this was so historically \cite{Penrose2010}) that the occurrence of the singularity sketched above may be an artifact of the high degree of symmetry of the solution being used. Quite the opposite, these features are completely generic, as was first understood by means of the so-called Penrose singularity theorem. This was the first of a sequence of results, usually known as Hawking-Penrose theorems (see \cite{Senovilla2014} for a recent account of the history of the subject and later developments up to date). These theorems are formulated on spacetime manifolds that possess a non-compact Cauchy hypersurface, i.e., the topology of the manifold is $\mathbb{R}\times\Sigma$ with $\Sigma$ being a non-compact spacelike tridimensional manifold. Roughly speaking, this means that the geometry does not incorporate new regions that require additional data for their description, and that there are well-defined notions of future and past.

The first notion we need concerns the ``surface'' of the black hole, that is, the two-dimensional manifold defined by the equation $r=r_{\rm s}$ at a given moment of time (being the Schwarzschild metric independent of time, this surface is stationary under its flow). We shall use the conventions and notation of \cite{Senovilla2014}. Let us imagine that we are emitting a spherical wavefront of light rays from a given position $r> r_{\rm s}$. Then light rays pointing outwards will move outwards, meaning that the radius of the corresponding sphere of light will grow in time, while light rays pointing inwards will move inwards, with decreasing radius in time. This is no longer true when we enter the Schwarzschild radius: for $r< r_{\rm s}$ both spheres of light move inwards. These surfaces are thus a particular case of the general concept of a closed future-trapped surface: a closed surface whose \emph{future-directed} null geodesics do not flow outwards. A closed past-trapped surface would correspond to the situation in which \emph{past-directed} null geodesics do not flow outwards. In the spherically symmetric case one can consider these surfaces as spherical, filling up the black hole as an onion. In this case the $r=r_{\rm s}$ surface marks the boundary between this trapping behavior and the normal behavior for $r>r_{\rm s}$, and represents the so-called trapping horizon, or marginally future-trapped surface.\footnote{In the Schwarzschild solution the event and trapping horizons coincide. This is not generally true for dynamical spacetimes, being the simplest example a small, non-spherical perturbation of the Schwarzschild solution \cite{Hawking1973}.}

The second notion is that of a singularity. This is a slippery concept as, strictly speaking, singularities do not belong to spacetime. We can however characterize the singular behavior by means of particles traveling to or from the singularity or, in geometric terms, of geodesics which finish or start at the singularity. In general, curves are maps from a subset of the real line $\mathbb{R}$ onto the spacetime manifold. When all the geodesics in a manifold are defined for the entire real line $\mathbb{R}$ when parametrized in terms of their affine parameter, the manifold is said to be geodesically complete, being incomplete in the opposite. The usual definition is that a spacetime manifold is singular if and only if it is geodesically incomplete.\footnote{Most definitions also include the condition of being non-extendable \cite{Hawking1973}, but this technical point is not really relevant for our purposes.} From a physical perspective this would imply the existence of particles or observers that disappear (or materialize) abruptly, which clearly corresponds to an abhorrent behavior. It is important to keep in mind that, for a manifold to be geodesically incomplete, it is enough that there exists just one incomplete geodesic.

The last ingredient is a condition on the curvature of the manifold. This crucial geometrical condition for the theorem is motivated by its translation, through the use of the Einstein field equations, into a condition upon the matter content. The so-called null convergence condition holds by definition if, for every null vector field $u^a$ in the manifold,
\begin{equation}
\mathscr{R}_{ab}u^au^b\geq0\label{eq:null1}
\end{equation}
with $\mathscr{R}_{ab}$ is the Ricci curvature tensor. If we apply the Einstein field equations,
\begin{equation}
G_{ab}=\mathscr{R}_{ab}-\frac{1}{2}\mathscr{R}g_{ab}=\frac{8\pi G}{c^4}T_{ab},\label{eq:einsteineqs}
\end{equation}
then Eq. \eqref{eq:null1} implies the null energy condition for the matter stress-energy tensor $T_{ab}$:
\begin{equation}
T_{ab}u^au^b\geq0.\label{eq:null2}
\end{equation}
With these ingredients, the Penrose singularity theorem is formulated as follows \cite{Penrose1965}: given a spacetime containing a non-compact Cauchy hypersurface $\Sigma$ and a closed future-trapped surface, the convergence condition \eqref{eq:null1} being valid for all the null vectors $u^a$ implies that there exist null geodesics that are incomplete in the future. 

In all its generality, this theorem does not provide us with any clue about the behavior of the spacetime near the singularity. To put an example, it does not assert that there is some curvature invariant (a scalar made up from the metric and its derivatives only) which blows up in the singularity, or that an entire set of observers in a given region (e.g., the interior of the future-trapped surface) will certainly experience the singular behavior in the future. Developing new techniques is necessary in order to tackle this thorny problem and understand in fine detail the physical properties of singularities. Among the attempts of doing so in the framework of the classical theory, the so-called BKL conjecture \cite{Khalatnikov1970,Belinskii1971}, contemporaneous to the singularity theorem stated above, occupies a prominent place. While in ordinary situations gravity is the weakest of all the forces in nature, singularities are the place in which gravity becomes the most prominent actor. Following this idea, in this conjecture it is assumed that matter fields do not play any essential role near the singularity. Moreover, it is postulated that the temporal derivatives of the metric field are the dominant ones, so that spatial derivatives can be ignored near the singularity (which is assumed to be spacelike). Every spatial point of spacetime will then become disconnected from the others, and will evolve following ordinary differential equations that present chaotic behavior. Numerical studies have shown evidence in favor of this behavior; see \cite{Berger2002,Ashtekar2011} for modern discussions.

\section{Black holes and spacetime singularities: ultraviolet effects beyond GR\label{sec:3}}

\subsection{Trapping horizons \label{sec:3a}}

When trying to build coherent scenarios in which the collapse of matter is affected by the very quantum nature of matter, the classical picture significantly changes. Robust semiclassical calculations tell us that black holes cannot be absolutely stationary, as they evaporate through the emission of Hawking quanta~\cite{Hawking1975,Unruh1976}. This is the last piece composing the so-called black-hole thermodynamics: a black hole of mass $M$ has a temperature
\begin{equation}
T_{\rm H}=\frac{\hbar c^3}{8\pi G M k_{\rm B}}.
\end{equation}
Given this result, it is assumed that the classical spacetime representing the collapse of a star to form a black hole should be modified so that the event horizon, instead of setting at a stationary position (once the absorption of surrounding matter and its stabilization has ended up), would shrink up to eventually disappear in a final explosion~\cite{Hawking1974}. It is also clear that the quantum effects responsible of the evaporation of the horizon would entail even more drastic modifications in the region surrounding the singularity. These realizations inevitably trigger the questions: Are these evaporating black holes really black holes in the sense of having an event horizon? Is there some information loss in a complete evaporation process? These questions have been a matter of controversy, and a strong driving force for theoretical development, since the publication of Ref. \cite{Hawking1976}. However, nowadays even Hawking concedes \cite{Hawking2014} that the most reasonable solution is that no event horizon would ever form, but only a structure which is locally similar but that lives for a finite, albeit extremely long, amount of time; that is, a trapping horizon as defined in Sec. \ref{sec:2b}. Under this view black holes, understood as causally disconnected regions of spacetime, would not strictly exist in nature. However, owing to the similarity of these quantum-corrected objects with classical black holes, they usually keep their name. Here we will generically call them regularly evaporating black holes (REBHs) to distinguish them from their classical cousins.  

\subsection{Preventing singularities \label{sec:3b}}

As happens with the event horizon the introduction of quantum mechanics, or any other kind of ultraviolet modifications, is likely to bring important changes to the classical picture concerning the singular behavior of GR. The first indication of these changes is that all kinds of energy conditions, that play a central role in the singularity theorems, are violated by quantum effects \cite{Barcelo2002,Ford2003}. Indeed, from a qualitative perspective it is widely expected that any consistent ultraviolet completion of GR will regularize the singular classical behavior. One of the ways to do so is the occurrence of repulsive forces preventing the gravitational collapse process to proceed indefinitely.\footnote{It would not be the first known example of this phenomenon: for example, the electron degeneracy pressure is a genuinely quantum effect that stabilizes white dwarfs \cite{Fowler1926,Chandrasekhar1931}.} Given their repulsive character, these effective forces will generally violate the energy conditions and, eventually, would unlock new options for the fate of gravitational collapse.

These qualitative expectations have been embedded in mathematical frameworks of different nature, essentially since the first explorations in semiclassical gravity \cite{Parker1973}, and most frequently in the field of cosmology \cite{Novello2008}. One of the most popular implementations, probably because of its fundamental flavor, is the one that results from the application of the loop quantum gravity techniques (see \cite{Ashtekar2005b} for an introduction to the subject). Loop quantum cosmology (see, e.g., \cite{Banerjee2012}) is the result of applying non-perturbative quantization techniques borrowed from the general theory and applied to some highly symmetric cosmological spacetimes. Although the results of this procedure have to be taken with a grain of salt \cite{Kuchar1989,Barbero2010}, these models are regarded as valuable tools in understanding the implications of the wider quantization program of the canonical structure of GR. One of the robust results of this approximation is that near the cosmological singularities, there appear effective forces with a net repulsive effect. These forces are strong enough to overcome the fatal attraction of gravity which would otherwise engender a singularity, provoking the \emph{bounce} of the matter distribution and connecting a classical contracting cosmology with a classical expanding cosmology \cite{Bojowald2001}; the Big Bang event is identified with the moment of the bounce, so that the expanding branch corresponds to our present universe. The bounce generally takes place when the density of matter is of the order of the Planck density,
\begin{equation}
\rho_{\rm c}\sim\rho_{\rm P}=\frac{m_{\rm P}}{\ell_{\rm P}^3}=\frac{c^5}{\hbar G^2}\simeq 5\times 10^{96}\mbox{ kg m}^{-3}.\label{eq:planckden}
\end{equation}
Following this result, quantum effects would be triggered when the radius of stellar structure is several orders of magnitude greater than the Planck length; for a star of mass $M$, its minimum radius $R_0$ will be approximately given by
\begin{equation}
R_0\sim\ell_{\rm P}\left(\frac{M}{m_{\rm P}}\right)^{1/3}.\label{eq:minrad}
\end{equation}
Although this may seem surprising at first, given that quantum gravity effects are usually bound to the Planck length, this very association is only reasonable in the absence of matter, that is, in the pure gravity case. Let us stress that the order of magnitude \eqref{eq:minrad} is the natural one on the basis of the Einstein field equations and the expectation that quantum gravity effects should appear when the curvature is of the order of $\ell_{\rm P}^{-2}$ \emph{in the presence of matter}, being independent of the specific ultraviolet completion one pursues for the gravitational interaction. Actually, for dust matter content of Planck density the trace of the Einstein field equations \eqref{eq:einsteineqs} leads to the very same order the magnitude:
\begin{equation}
\mathscr{R}=g^{ab}\mathscr{R}_{ab}=-\frac{8\pi G}{c^4}g^{ab}T_{ab}=\frac{8\pi G\rho_{\rm P}}{c^2}=\frac{8\pi}{\ell_{\rm P}^2}.\label{eq:planckcur}
\end{equation}
Thus in the presence of matter the association between Planckian densities and ultraviolet effects is completely straightforward, independently of the specific ultraviolet completion one considers.

In loop quantum cosmology, the bouncing behavior is successfully captured by an affective Friedmann equation \cite{Taveras2008} which, in the flat ($k=0$) case, takes the form 
\begin{equation}
\left(\frac{\dot{a}}{a}\right)^2=\frac{8\pi G}{3}\rho\left(1-\frac{\rho}{\rho_{\rm c}}\right).\label{eq:frimod}
\end{equation}
In this equation it is easy to see that reaching the critical density corresponds to a turning point, $\dot{a}=0$. The second term in the right-hand side of this equation corresponds to an effective stress-energy tensor that strongly modifies the classical behavior. It will be instructive to have in mind this feature in our later discussion of gravitational collapse. The occurrence of effective repulsive forces will probably be a robust ultraviolet effect, independently of the specific quantization program or ultraviolet completion one choses (see, e.g., \cite{Barcelo2011,Parker1973,Broda2011,Abedi2015} for different suggestions), leading eventually to a resolution of the singular behavior of the classical theory. However, while the cosmological situation is quite well understood nowadays, how these considerations would affect the issue of gravitational collapse is a more subtle issue, as we discuss in the following.

\section{Ultraviolet effects: phenomenological considerations\label{sec:4}}

\subsection{Horizon predominance \label{sec:4a}}

The necessity of finding an ultraviolet completion of GR has been in the air at least since the first successes of quantum field theory, in particular when applied to electrodynamics. The generalizations of these field-theoretical developments to GR were plagued with difficulties, thus forcing to use partial, and therefore necessarily incomplete, approaches. These have been used since the sixties to get insights into the ultraviolet behavior of gravitational collapse processes. Before Hawking's groundbreaking developments, there already were some interesting attempts to see what could result from the incorporation of quantum-mechanical effects to a black hole spacetime. As early as 1966, Sakharov~\cite{Sakharov1966} and Gliner~\cite{Gliner1966} suggested that at the high densities close to singularity formation the effective matter content might develop a vacuum-like equation of state $\rho=-p$, providing a repulsive force. Soon after, Bardeen showed that it is possible to construct black hole geometries (here meaning having an event horizon) satisfying the null energy condition but having no singularities~(\cite{Bardeen1968}, see also~\cite{Borde1994}). Singularity avoidance (and hence a way out of the singularity theorems) is permitted due the non-existence of an open Cauchy surface: the spacetime develops a topology change from open hypersurfaces to closed ones~\cite{Borde1997}. Matter crossing the black-hole horizon and falling towards the apparent singularity would end up reappearing from a white-hole horizon into a \emph{different universe}, i.e., not in the same asymptotic region. Bardeen's spacetime and others with similar structure (e.g.~\cite{Dymnikova1992,Mars1996,Ayon-Beato1998,Bronnikov2001,Bronnikov2001b,Hayward2006,Olmo2014,Olmo2015}) are examples of regularizations of the black hole singularity that do not affect the event horizon.

Hawking's idea that event horizons should evaporate, leading eventually to some fundamental loss of information, introduced a new ingredient into the singularity problem. Many different evaporation scenarios have been put forward since then to accommodate solutions to the information and singularity problems. These proposals have generated a large amount of work; see for instance the sample of recent papers \cite{Ashtekar2006,Ashtekar2008,Hossenfelder2010,Kawai2013,Kawai2014,Torres2014,Frolov2014,Bojowald2014,Gambini2014,Torres2015,Rovelli2014,Barrau2014b,Bambi2013,Liu2014,Zhang2015,Modak2015,Modak2015b,Hawking2015,Hooft2015}. What we want to emphasize here is that, at least qualitatively, the resulting effective geometries advocated by most of these scenarios (what we have called REBHs) are essentially equivalent. Diagrams describing these geometries can be found for example in~\cite{Frolov1979,Frolov1981,Roman1983,Ashtekar2005,Hayward2006,Rovelli2014,Bambi2014,Bardeen2014,Kawai2015}. Aside from their fine theoretical details, all of them contemplate REBHs as essentially hollow and almost stationary regions of spacetime; all of them share the substitution of event horizons by \emph{extremely slowly} evaporating trapping horizons as seen by external observers, even if the specific time scales could vary between different models. In this sense these proposals may be regarded as ``conservative''.

Thus, the prevalent view is that REBHs indeed form in astrophysical scenarios and, ignoring their gravitational interaction with the surrounding matter, they would remain almost inert for very many Hubble times except for a tiny evaporative effect that would eventually make them disappear. The evaporation rate of stellar-mass objects would be so slow, $10^{67}$ years to halve the size of a Solar mass object, that for all practical purposes they could be considered stationary. REBHs have a ridiculously large lifetime, in whatever measure: the estimated age of the universe is of the order of $10^{10}\mbox{ years}$. Whereas the distinction between a classical BH and a REBH is of fundamental interest on purely theoretical grounds, it is almost certainly irrelevant for all astrophysical purposes, and arguably for any meaningful operational perspective \cite{Visser2014}. This becomes even worse when considering that astrophysical black holes are actually on average growing and not yet evaporating because their Hawking temperature is smaller than the approximately 3K of the cosmic microwave background \cite{Samtleben2007}. This situation could be partially alleviated if primordial black holes were generated in our universe, as it was already proposed by Hawking~\cite{Hawking1974}; see~\cite{Carr2010} for some recent constraints on their existence. In this paper we shall focus on contemporary gravitational collapse processes, which are certainly the most interesting ones from an experimental perspective.

Given the irrelevance of Hawking radiation for astrophysical purposes, this implies that any experimental test, whose result is available to observers outside black holes, of the precise way in which the evaporation proceeds would be almost certainly beyond the reach of humankind. Most importantly, the very nature of any kind of process taking place in the interior of the trapping horizon, and in particular the fate of the matter inside it and the ultraviolet regularization of the singularity, will be inevitably hidden for us unless something radical changes this picture. While observers that fall into the black hole and cross the trapping horizon will be able to catch a glimpse of what is going on inside, they will not be able to communicate their experience to the exterior, enforcing the ignorance of external observers.\footnote{We encourage the reader to note the amusing analogy with the existence of life after death.}

For the standard REBH perspective the only possible, albeit partial, experimental corroboration of the overall scenario would be the real detection of Hawking radiation.\footnote{Even this assertion is debatable, as stellar objects hovering near its Schwarzschild radius could produce a radiation signal very close to the predicted Hawking radiation \cite{Stephens1994,Barcelo2006,Barbado2011}.} In the real world the systems hosting black holes are so complex that there exist multitude of forms of radiation that would eventually frustrate any attempt to measure such a tiny effect.\footnote{A great deal of effort has been put on the detection of the analogue of the Hawking effect in the analogue gravity framework \cite{Barcelo2005,Steinhauer2014,Steinhauer2015}. It is, however, unclear to what extent the analogy is deep enough to consider the measure of these effects on fluids a detection of the `genuine' gravitational Hawking radiation; some arguments in favor can be read in \cite{Unruh2014}.} On the other hand, given the ridiculously long times associated with the evaporation process, any event happening inside the trapping horizons formed in gravitational collapse processes is certainly irrelevant for practical purposes. Any attempt to understand the physics behind the horizon, for instance on the lines sketched in Sec. \ref{sec:3b}, would be unquestionably pointless: while the fate of matter behind the trapping horizon and the horizon itself could prove an interesting intellectual exercise, it would be impossible to find any experimental corroboration of these developments. The black hole trapping horizon will effectively act as an event horizon, forcing our ignorance about what is behind. This observation is not a matter of improving the sensitivity of experiments, but the issue is a completely different (and for us, quite unpleasant) one.

One could argue that, even accepting that the gravitational collapse process itself could be useless to boost our understanding of the physics near singularities, the Big Bang singularity could be used as a substitute from which to obtain this knowledge. The first observation that comes to mind concerning this assertion is the issue of repeatability: while we expect that numerous processes of gravitational collapse are taking place now around us in the universe, and will be quite surely taking place in the future, in the currently accepted model of the universe there is only one Big Bang event. Also the Big Bang singularity lies in the past so that, while we can hopefully describe its properties in a simple way, it is questionable to what extent we are able to perform experiments in the usual sense of the word, or if the most we can hope to do is constructing some sort of cosmological ``archeology''. Leaving aside these issues, which are indeed ubiquitous to the field of cosmology (see \cite{Goenner2010} for instance), there are good physical reasons to believe that the singularities associated with black holes should be different from the initial cosmological singularity. These arguments are based on the second law of thermodynamics and have been repeatedly exposed by Penrose \cite{Penrose2010}. While one may think that these singularities correspond from a mechanical point of view as the time-reversal of each other, thermodynamical considerations break this apparent time reversal. On the one hand, if the second law of thermodynamics is to be applicable to the universe, the behavior near the initial singularity is to be associated with a low amount of entropy. On the other hand, the process of gravitational collapse to black holes is generally expected to lead to high entropies. In particular, this should indeed be the case in the standard scenario, in order to fulfill the thermodynamic description of black holes and match the Bekenstein-Hawking entropy formula
\begin{equation}
S=\frac{c^3 A}{4 G\hbar},
\end{equation}
where $A$ is the area of the horizon. This points to a deep physical distinction between these situations that supersedes the apparent time reversal symmetry.

To end this section let us say a few words about other popular and somewhat more exotic approaches to the information and singularity problems. Our comments here will be restricted to their relation with the lifetime problem which is central for this work. The complementarity approach of Susskind~\cite{Susskind1993} would share the REBH geometric description but maintaining that it would be only relevant for observers actually crossing the horizon. A complete experimental proof of the overall complementarity scenario would therefore face the same practical problems than more standard REBHs scenarios. The fuzzball proposal of Mathur~\cite{Mathur2005,Mathur2009} takes the view that the evaporating horizon marks a real border for the spacetime. Information is accumulated at the gravitational radius being retrieved only after enormous amounts of time. The same situation occurs in the condensed state scenario of Dvali~\cite{Dvali2014,Dvali2015}. Let us recall that the Page time (half-mass evaporation, the assumed time of retrieval of the information in these scenarios) for an astrophysical black hole is as extremely long as the time scale associated with Hawking evaporation. The Giddings remnant scenario~\cite{Giddings1992,Giddings1994} adds to the REBHs the presence of massive remnants. Only the last stages of the REBHs picture are expected to be modified so that this scenario also shares the lifetime problem. 

Let us note for completeness that there is a recent construction by Haggard and Rovelli~\cite{Haggard2014} that shares some geometric properties with our proposal, as it also describes the black to white hole transition with only one asymptotic region. However, it is completely different in essence: by construction, the duration of the total transition as seen from external observers is extremely long. Therefore this scenario is indistinguishable for an extremely long time from any REBH model, being therefore not better regarding the lifetime problem.

\subsection{Preponderance of the singularity regularization\label{sec:4b}}

The possibility of measuring the ultraviolet effects that are responsible for singularity avoidance in reasonable time scales would imply that Hawking radiation would not be able to carry away from the object a significant amount of energy in the meanwhile. Indeed, the horizon barrier should be broken in some way so that some signals from the inside may reach an external observer and, as the evaporation process is extremely slow, the net evaporative effect would be negligible. Even if we wait the entire age of the universe (specifically, the Hubble time $t_{\rm H}$), the energy loss of a stellar-mass black hole would be of the order of
\begin{equation}
10^{-4}\,\frac{\hbar c^6}{G^2}\frac{t_{\rm H}}{M^2_\odot}\sim10^{-11}\mbox{ J}.
\end{equation}
This represents a $10^{-49}$ part of its original rest energy, which is absurdly negligible. Again, from an operational perspective only part of these theoretical developments associated with gravitational collapse could be relevant. We end up with a dilemma: if the physics of horizons as long-lived thermodynamical entities is to be realized in nature, there is very little hope that we will be able to observe any trace of the physics associated with the corresponding (regulated) singularities while, if the physics of these spacetime singularities is to be observable in reasonable time scales, any dissipation through the emission of Hawking radiation would be physically irrelevant in energetic terms.

An additional tension, of a different kind but related to the previous one, appears when ultraviolet effects inside the collapsing distribution of matter are taken into account. As a first approximation let us consider the homogeneous and isotropic situation inside the star, so that we can use a patch of a cosmological solution to describe the internal geometry. In the pressureless case the surface of the star (as well as any point in the interior) will follow a geodesic in the external metric, i.e., the Schwarzschild metric. Any nonzero pressure will lead to deviations from this behavior. As an extreme example one could imagine a strong enough pressure to halt the collapse before the event horizon is formed; the radius of the star will then reach a fixed value. While this is what happens for structures below the Tolman-Oppenheimer-Volkoff limit of neutron stars \cite{Oppenheimer1939,Rhoades1974,Bombaci1996}, for larger masses there is no known force which can compete with the gravitational contraction \cite{Woosley2002}. It seems unavoidable that, in these situations, the matter distribution crosses its Schwarzschild radius, so that we can even take the idealized Oppenheimer-Snyder model, neglecting any known form of pressure with respect to the gravitational force. Deviations from GR would not appear until high enough densities are reached, and certainly \emph{not} at the moment when the star generates its trapping surface (at least if the collapse proceeds in a free-fall manner; departures from this behavior could lead to strong deviations even before of horizon crossing \cite{Barcelo2008}). As we discussed above, if singularities are to be avoided in a suitable ultraviolet completion, a universal effective pressure with a net repulsive effect should appear at some point. This pressure would prevent the formation of singularities, leading to bouncing solutions of similar nature as the ones that were previously described in a cosmological framework. How is this compatible with the geometry outside the star?

The bouncing process of the matter distribution implies that the radius of the star should grow in proper time for an observer in the surface. But this is in contradiction with the external metric: there are no timelike geodesics of non-decreasing radius inside the Schwarzschild radius. In order to avoid a causality violation we should accept a modification of the external metric around the bouncing star. Given the spherical symmetry we are assuming, these geometry deviations should take the form of an effective matter content violating energy conditions. In other words, these general arguments point to the necessary existence of modifications of the geometry even outside the star. These modifications are generally expected to be localized in a small region around the minimum radius of the star $R_0$ in which the curvature invariants associated with the classical geometry are still large enough.

The situation is, in our opinion, more subtle. Once we have accepted that a bounce of the star should occur near the singularity, both geometric pictures inside and outside the radius of the star seem to be hardly reconcilable. More specifically, gluing these geometries along the wordline (for fixed angular variables) of the surface of the star no longer appears as a viable option, as these present what we can regard as competitive trends. While the external geometry will try to stop the matter distribution to escape, the internal geometry of matter will try to overcome the overwhelming horizon influence. It seems that one should rather face the problem of how these effects compete in order to have a glance to the final outcome. In the most general case the resulting geometry will present a transition layer between these two regimes. How to describe the properties of this layer seems to be currently unknown.

In the standard view of REBHs, the internal geometry is completely dominated by the external geometry, so that the transition layer affects the behavior of the overall internal metric. This would correspond, e.g., to the effective stabilization of the bouncing star at some asymptotic radius \emph{inside} the horizon. Such a situation corresponds to one of the extremes of the entire family of possible interpolating geometries. The situation we want to analyze here is the complementary extreme, with the internal geometry unchanged and the external metric modified in order to guarantee a smooth matching. As we will discuss, in this case the transient may be analogous to a shock wave phenomenon as those occurring in normal fluids.

From a theoretical perspective, the fine knowledge of the resulting geometry will probably have to wait until our understanding of the ultraviolet properties of the gravitational interaction is deep and firm enough to tackle these complex questions. Although ultraviolet effects are expected to be triggered only when curvatures are high enough (i.e., Planckian), we have virtually no knowledge of how these regions of high-curvature, that violate energy conditions, evolve once they are generated, how they propagate and decay in time. While a first-principles study of this issue is highly interesting, we can always revert the logic and try a phenomenological approach. We can study the possible effective geometries and try to extract physical and observational consequences, information that could be eventually used in order to progress in our theoretical understanding.

\section{Effective bounces, black to white hole transitions and shock waves \label{sec:5}}

In plain terms, what we want to analyze is what would happen if the spherically-symmetric collapsing matter, upon reaching Planck density, slowed down and bounced back. In geometric terms we expect the description near the matter crossing the horizon outwards to correspond with the time-reversed geometry to that of a black hole, namely the one associated with a white hole. Birkhoff's theorem is certainly a uniqueness result for the vacuum geometry outside a spherical distribution of matter with mass parameter $M$ and radius greater than its Schwarzschild radius. However, for vacuum geometries inside the Schwarzschild radius this is not true. Birkhoff's theorem asserts that any vacuum patch must be locally isometric to a patch of the maximally extended Kruskal manifold~\cite{Hawking1973}. The vacuum geometry outside a spherical distribution of matter but still inside the Schwarzschild radius could therefore either correspond to the black-hole or the white-hole patch. From this perspective, the bounce can be represented by a transition between these two patches which will necessarily contain features that go beyond GR. 

It is an essential feature that the flat asymptotic regions corresponding to these geometries are one and the same. The bounce is a physical event taking place entirely in our universe; it does not describe the escape of matter to remote regions that are causally disconnected from the static observers standing outside the stellar structure (as discussed in Sec. \ref{sec:4a}, proposals of this kind fall again within the REBH paradigm). As we will see later, the most general metric describing this bounce contains a number of unknown parameters, so that its fine details certainly depend on the underlying ultraviolet completion of GR. However, it also possesses some robust characteristics that clearly distinguishes it from REBH proposals even from a physical perspective. Let us proceed in a constructive way, and build first a simple toy-model geometry of a time-symmetric bouncing regularization of a collapsing star, that however encapsulates the most relevant features of the process we want to describe. While  this geometry will present some singular properties, it can be considered as the distributional limit of well-behaved geometries, thus providing indeed a very good analytical approximation.

We shall describe the bouncing geometry from an explicitly time-symmetric point of view. As stated above, for simplicity we are concentrating on the spherically symmetric collapse, with the collapsing matter being characterized as a pressureless dust cloud. Also we will ignore dissipation effects in a first stance: classically, a spherically-symmetric collapse does not produce any dissipation in the form of gravitational radiation. If we took into account quantum corrections, there would indeed exist some (rather small) dissipation. In the limit of very large masses, this quantum radiation could in principle be made as small as desired. The time-symmetric geometry we are going to present is a reasonable dissipationless approximation to more realistic, dissipative situations, that we explore in Sec. \ref{sec:6}.

It will prove useful to describe the metric using generalized Painlev\'e-Gullstrand coordinates~\cite{Kanai2011}. These coordinates are adapted to observers attached to the stellar body. We shall write the metric as 
\begin{eqnarray}
ds^2=-A^2(t,r) dt^2 + {1 \over B^2(t,r)}[dr - v(t,r) dt]^2+r^2 d\Omega^{\,2}_2,
\label{generalized-acoustic-metric}
\end{eqnarray}
with $d\Omega^{\,2}_2$ being the line element of the unit 2-sphere. The three functions $A(t,r)$, $B(t,r)$ and $v(t,r)$ will in the following be given patch by patch. Let us begin considering that the collapse process starts at rest at infinity, so that the line element \eqref{generalized-acoustic-metric} reduces to the the standard Painlev\'e-Gullstrand one, with $A^2(t,r)=c^2$ and $B^2(t,r)=1$. When familiarized with this situation, we will move on to consider that the more general case in which the collapse starts from an initial radius $r_{\rm i}$ in Sec. \ref{sec:5c}.

\subsection{Collapse from infinity and homogenous thin-layer transition \label{sec:5a}}

To describe the bounce, we shall glue two geometries, one corresponding to the Oppenheimer-Snyder collapse of a homogeneous ball of dust, the other one being its time-reversal. Let us consider for the moment the metric outside the star, corresponding to the Schwarzschild solution. The velocity profile $v(t,r)$ presents a flip of sign between the black-hole and white-hole patches of the Kruskal manifold. Thus the metric we are seeking for will be characterized by a velocity profile
\begin{equation}
v(t,r) =v_{\rm s}(r)[1-2\theta(t)]=-c\,[1-2\theta(t)]\sqrt{{r_{\rm s}\over r}},\qquad R(t)\leq r\leq+\infty. 
\label{velocity-profile_1}
\end{equation}
Here $v_{\rm s}(r)=c\sqrt{r_{\rm s}/r}$ is the absolute value of the standard velocity profile of the Schwarzschild solution, and the Heaviside function $\theta(t)$ is used to perform the transition between the different patches. Concerning the function $R(t)$ describing the trajectory of the surface of the star, for the moment the only condition we demand, following our previous discussion, is that it is bounded from below by $R(0)=R_0$ where $t=0$ corresponds to the moment of the bounce in the coordinates we are using.

This gluing procedure by itself is nothing but a brute force exercise, and the result presents unpleasant features. However, we want to present a constructive procedure in which the details are progressively added on demand, eventually arriving to the complete picture in all its generality. For instance, using the Heaviside function to construct the geometry makes all the $t=0$ hyperplane singular for $r>R_0$, in the sense that the metric is discontinuous there. Of course, this does not necessarily reflect any physical reality, as we can always introduce a regulator to make the geometry finite. Let us introduce the continuous, differentiable but non-analytic function
\begin{equation}
f(t)=\left\{\begin{array}{cr}\exp(-1/t)\qquad&0\leq t,\\ 0\qquad&t\leq0,\\\end{array} \right.
\end{equation}
that permits us to define for $t_{\rm R}\in\mathbb{R}$ the function
\begin{equation}
g_{t_{\rm R}}(t)=\frac{f(1/2+t/t_{\rm R})}{f(1/2+t/t_{\rm R})+f(1/2-t/t_{\rm R})}.\label{eq:interfun}
\end{equation}
This function represents a regulated version of $\theta(t)$ that interpolates between the values $g_{t_{\rm R}}=0$ (for $t\leq -t_{\rm R}/2$) and $g_{t_{\rm R}}=1$ (for $t\geq t_{\rm R}/2$). The value of $t_{\rm R}$ thus controls the duration of the interpolation. We can actually define the pointwise limit
\begin{equation}
\theta(t)=\lim_{t_{\rm R}\rightarrow 0}g_{t_{\rm R}}(t),
\end{equation}
thus fixing the convention $\theta(0)=1/2$.

\begin{figure}[h]%
\vbox{ \hfil  \includegraphics[width=0.55\textwidth]{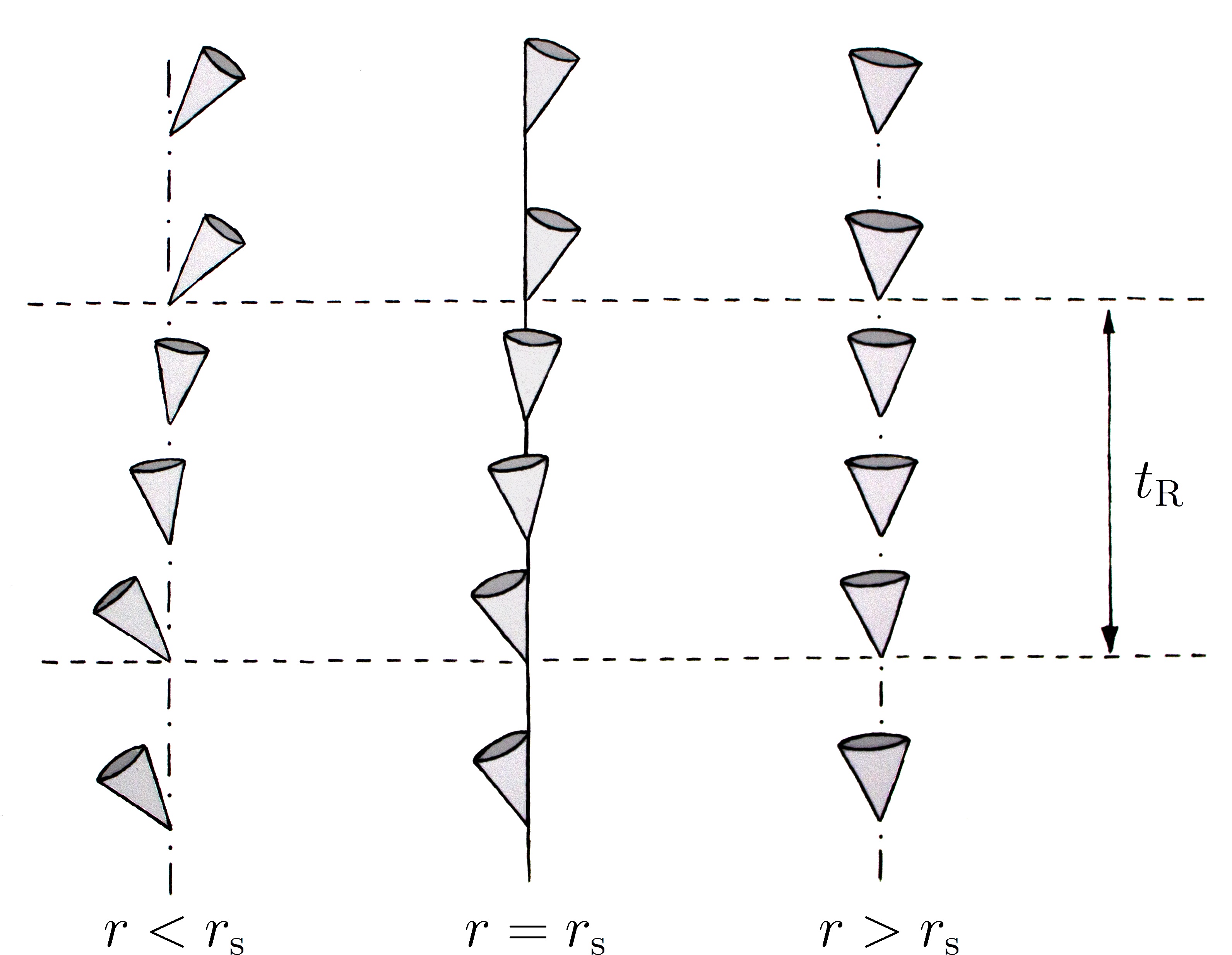}\hfil}
\bigskip%
\caption{Transition from the black-hole patch to the white-hole patch by using a smooth interpolation with characteristic time scale $t_{\rm R}$. The qualitative behavior of light cones for different values of the radial coordinate is shown.}
\label{Fig:transition}%
\end{figure}%

With respect to the field equations, the strict limit $t_{\rm R}\rightarrow 0$ may not well defined as a solution even in the distributional sense, as its curvature may contain terms that are quadratic in the Dirac delta function, which are ill-defined (in general, the product of two distributions cannot be defined \cite{Grubb2009}). Let us emphasize again that we always keep in the back of our minds that the relevant physical situation corresponds to $t_{\rm R}$ small, but nonzero. For $t_{\rm R}>0$ we can safely insert the metric in the Einstein field equations \eqref{eq:einsteineqs} to obtain (using, e.g., Eqs. 3.2 in \cite{Kanai2011}) the following nonzero components of the effective stress-energy tensor supporting this geometry:
\begin{align}
\mathscr{T}^1_{\,\,\,\,\,0}=-\frac{c}{2\pi}(1-2g_{t_{\rm R}})\dot{g}_{t_{\rm R}}\frac{M}{r^2},\nonumber\\
\mathscr{T}^1_{\,\,\,\,\,1}=-\frac{c}{2\pi}\dot{g}_{t_{\rm R}}\frac{M}{r^2}\left(\frac{r_{\rm s}}{r}\right)^{-1/2},\nonumber\\
\mathscr{T}^2_{\,\,\,\,\,2}=-\frac{c}{8\pi}\dot{g}_{t_{\rm R}}\frac{M}{r^2}\left(\frac{r_{\rm s}}{r}\right)^{-1/2}.
\label{eq:effset}
\end{align}
We use the notation $\mathscr{T}_{ab}$ in order to make a clear distinction between this effective source originated in the bounce, and the stress-energy tensor $T_{ab}$ of the perfect fluid that forms the stellar structure. It is interesting to note that the effective stress-energy tensor \eqref{eq:effset} only involves first derivatives of the interpolating function $g_{t_{\rm R}}$, so that the distributional limit is indeed well-defined, corresponding to a $\delta-$function source located on $t=0$; for $t_{\rm R}>0$ this source presents certain width. The effective source \eqref{eq:effset} violates the null energy condition \eqref{eq:null2} in all its radial extension; let us consider for example the hyperplane $t=0$ and the null vector $u^a$ on it, with components $u^0=u^1=1$, $u^2=u^3=0$. Then
\begin{equation}
\mathscr{T}_{ab}u^au^b=-\frac{c}{\pi}\frac{M}{r^2}\left(\frac{r_{\rm s}}{r}\right)^{-1/2}\left.\dot{g}_{t_{\rm R}}\right|_{t=0}<0.
\end{equation}
Taking into account the metric inside the collapsing star, the complete geometry is given by:
\begin{equation}
A^2(t,r)= c^2 B^2(t,r)=c^2,
\end{equation}
and
\begin{eqnarray}
v(t,r) =-c\,[1-2g_{t_{\rm R}}(t)]\times\left\{
\begin{array}{cr}
\displaystyle\sqrt{{r_{\rm s} \over r}}\qquad &R(t)\leq r, \\
\displaystyle{r \over R(t)}\sqrt{{r_{\rm s} \over R(t)}}\qquad &r\leq R(t). 
\end{array}
\right.
\label{velocity-profile1}
\end{eqnarray}
The function $R(t)$ is constructed from the calculation in GR of the Oppenheimer-Snyder collapse starting from infinity \cite{Kanai2011}:
\begin{eqnarray}
R(t)=\left(\frac{9r_{\rm s}}{2}\right)^{1/3}|c\,t|^{2/3}.
\end{eqnarray}
The point $t=0$ in this trajectory would correspond to the classical singularity. For $t<0$ the trajectory goes inwards, while for $t>0$ it goes outwards. To make the geometry smooth, we shall consider that the function $R(t)$ is modified by means of a smooth interpolation in the interval $t\in[-t_{\rm R}/2,t_{\rm R}/2]$ so that the classical singularity is never formed; as said before, $R(t)$ should be bounded from below by $R(0)=R_0>0$. The order of magnitude of $R_0$ is fixed through model-independent considerations as given by Eq. \eqref{eq:minrad}.

To sum up, the two main features of the geometry presented in this section are the regularization of the singularity by the bounce of the collapsing matter at $R(0)=R_0$ and the existence of a thin-layer transition region between the black-hole and white-hole patches that violate the null convergence condition \eqref{eq:null1}. This thin layer is homogeneous in the sense that the interpolation time $t_{\rm R}$ is independent of the radial coordinate; this dependence will be included in Sec. \ref{sec:5c}. Let us discuss for the moment some additional relevant properties of this region of spacetime.

\subsection{Non-perturbative ultraviolet effects \label{sec:5b}}

The curvature scalar corresponding to the effective source \eqref{eq:effset} is given by
\begin{equation}
\mathscr{R}=6\frac{\dot{g}_{t_{\rm R}}}{c}\sqrt{\frac{r_{\rm s}}{r^3}}.\label{eq:infcur}
\end{equation}
This expression permits us to make an order of magnitude estimate for $t_{\rm R}$, by the following argument. Let us assume that, when the stellar radius reaches its minimum value, $R_0$, its density is of the order of the Planck density, corresponding to Planckian curvatures by virtue of \eqref{eq:planckcur}. These Planckian curvatures generate non-perturbative contributions that prevent the structure from collapsing, triggering the bounce process [recall the form of the modified Friedmann equation \eqref{eq:frimod}]. For continuity reasons, it is reasonable to expect that the modifications of the classical geometry outside, but very close to the stellar radius, are at this stage also Planckian. This is guaranteed if, as using Eqs. \eqref{eq:minrad} and \eqref{eq:infcur}, and the estimation $\dot{g}_{t_{\rm R}}\sim 1/t_{\rm R}$ reveals,
\begin{equation}
t_{\rm R}\sim \frac{\ell_{\rm P}}{c}=t_{\rm P},\label{eq:treg}
\end{equation}
with $t_P$ the Planck time. These extremely short times, in conjunction with the mechanism acting as the trigger and the information about the curvature \eqref{eq:infcur}, suggest the following interpretation of the effective source \eqref{eq:effset} by analogy: it may be understood as a shock wave, originated by the violent bounce of the stellar structure at $r=R_0$, that propagates outwards. The decay rate of this ``curvature wave'' with the radius goes as $r^{-3/2}$, which is to be compared with, e.g., the decay of the classical value of the Kretschmann scalar $\mathscr{K}=\mathscr{R}_{abcd}\mathscr{R}^{abcd}$ of the Schwarzschild geometry, valid for $|t|>t_{\rm R}/2$,
\begin{equation}
\sqrt{\left.\mathscr{K}\right|_{|t|\geq t_{\rm R}/2}}=2\sqrt{3}\,\frac{r_{\rm s}}{r^3}.\label{eq:kre1}
\end{equation}
Indeed, to be more consistent we may compare this expression with the ultraviolet corrections to the Kretschmann scalar at $t=0$ for instance. The evaluation of the leading order of this quantity for short transients shows that it nevertheless behaves as Eq. \eqref{eq:infcur}:
\begin{equation}
\sqrt{\left. \Delta \mathscr{K}\right|_{t=0}}\simeq 6\sqrt{2}\,\frac{\dot{g}_{t_{\rm R}}}{c}\sqrt{\frac{r_{\rm s}}{r^{3}}}.\label{eq:kremod}
\end{equation}
While on $R_0$ the classical and ``ultraviolet'' parts of the Kretschmann scalar are of the same order by construction, their decay rates for $r>R_0$ are rather different. A look at Eqs. \eqref{eq:kre1} and \eqref{eq:kremod} suffices to check that the effects associated with the shock wave decrease more slowly than the curvature of the classical geometry. We could say that the wave preserves its non-perturbative character, with respect to the background, in its way out. As we will discuss in detail in Sec. \ref{sec:5f}, this is the mechanism that permits to understand how ultraviolet modifications are able to alter dramatically the geometry on the horizon, even if the curvatures might no longer be Planckian there. Notice that the infinite radial extension of this shock wave is an unphysical feature of this simplified model that will not survive in more elaborated models such as we explain below.

\subsection{Collapse from a finite radius and triangular-shaped transition \label{sec:5c}}

Up to now, we have considered that the initial radius of the stellar structure is infinite. Let us look for the description of the more realistic case of gravitational collapse from a finite radius $r_{\rm i}$. The resulting metric will be written again in the form of Eq. \eqref{generalized-acoustic-metric}. The collapse of the star begins at rest at the initial radius $r_{\rm i}$ and is represented by the trajectory of the star's surface $R(t)$. The arbitrary zero of time is chosen such that the collapse 
starts at $t=-t_{\rm B}/2$, where
\begin{equation}
t_{\rm B}=\pi\, \frac{r_{\rm i}}{c}\sqrt{\frac{r_{\rm i}}{r_{\rm s}}}\label{eq:tbounce}
\end{equation}
is twice the classical collapsing time from $r=r_{\rm i}$ to $r=0$ in the Oppenheimer-Snyder model. For $t\leq-t_{\rm B}/2$, i.e., before the collapse has started, the metric is exactly Schwarzschild outside $r_{\rm i}$, while for $r\leq r_{\rm i}$, 
\begin{equation}
A^2(t,r)=c^2,\qquad B^2(t,r)=1-\frac{r_{\rm s}r^2}{r_{\rm i}^3},\qquad v(t,r)=0.
\end{equation}
This metric corresponds to a star of homogeneous density, maintained static by an appropriate internal pressure. Once the collapse has started, the three coordinate patches we use are $0\leq r\leq R(t)$, $R(t)\leq r\leq r_{\rm i}$, and $r_{\rm i}\leq r\leq +\infty$. The functions describing the metric are then:
\begin{eqnarray}
A^2(t,r)=c^2\times\left\{
\begin{array}{cr}
\displaystyle \frac{1-r_{\rm s}/ r}{1-r_{\rm s}/r_{\rm i}}\qquad &r_{\rm 
i}\leq r, \\
1\qquad &R(t)\leq r\leq r_{\rm i}, \\
1\qquad &r\leq R(t); 
\end{array}
\right.
\label{c-profile}
\end{eqnarray}
\begin{eqnarray}
B^2(t,r)= \left\{
\begin{array}{cr}
\displaystyle1-r_{\rm s} /r\qquad &r_{\rm i}\leq r, \\
\displaystyle1-r_{\rm s}/r_{\rm i}\qquad &R(t)\leq r\leq r_{\rm i}, \\
\displaystyle1-{r_{\rm s} \over r_{\rm i}}\left({r \over R(t)}\right)^2\qquad &r\leq R(t);
\end{array}
\right.
\label{epsilon-profile}
\end{eqnarray}
\begin{eqnarray}
v(t,r)=-c\,[1-2g_{t_{\rm R}(r)}(t)]\times\left\{
\begin{array}{cr}
0\qquad &r_{\rm i}\leq r, \\
\displaystyle\sqrt{{r_{\rm s} \over r}-{r_{\rm s} \over r_{\rm i}}}\qquad &R(t)\leq r\leq r_{\rm i}, \\
\displaystyle{r \over R(t)}\sqrt{{r_{\rm s} \over R(t)}-{r_{\rm s} \over r_{\rm i}}}\qquad &r\leq R(t). 
\end{array}
\right.
\label{velocity-profile}
\end{eqnarray}
In this case, the function $R(t)$ is constructed from the calculation in GR for the Oppenheimer-Snyder collapse from a finite radius $r_{\rm i}$:
\begin{eqnarray}
R(t)= {r_{\rm i} \over 2}(1 +\cos \eta),\qquad t= \frac{t_{\rm B}}{2\pi}(\eta +\sin \eta-\pi),\label{eq:oppsny}
\end{eqnarray}
with $\eta \in [0,2\pi]$. The point $\eta=\pi$ ($t=0$) in this trajectory would correspond to the classical singularity; for $\eta<\pi$ the trajectory goes inwards, while for $\eta>\pi$ it goes outwards. Again, we shall modify the function $R(t)$ so that it is bounded from below by $R(0)=R_0>0$. 

A new feature in this situation is the necessary introduction of a radial dependence on the parameter $t_{\rm R}$ that controls the interpolation time, which then becomes a function $t_{\rm R}(r)$. This quantity enters through the function $g_{t_{\rm R}(r)}(t)$ in Eq. \eqref{velocity-profile}. This is necessary to guarantee that curvature invariants are kept finite in the surroundings of the $r=r_{\rm i}$ hypersurface. Had we taken this parameter as being constant, the leading order of the Ricci scalar in the limit $r\rightarrow r_{\rm i}$ from below would be given, at $t=0$ for instance, by
\begin{equation}
\left.\mathscr{R}\right|_{t=0}\simeq \frac{4r_{\rm s}(3r_{\rm i}-4r)}{r_{\rm i}\,r^2\sqrt{\displaystyle\frac{r_{\rm s}}{r}-\frac{r_{\rm s}}{r_{\rm i}}}}\frac{1}{c\,t_{\rm R}}.\label{eq:domricc}
\end{equation}
On the other hand, the consideration of \emph{decreasing} functions $t_{\rm R}(r)$ with the radius is clearly motivated by the understanding of the modifications of the near-horizon geometry as the result of the propagation of non-perturbative ultraviolet effects from $r=R_0$ up to $r=r_{\rm i}$. It is therefore quite remarkable that avoiding that the scalar curvature blows up at $r=r_{\rm i}$ at $t=0$ demands that $t_{\rm R}(r)$ verifies the differential equation
\begin{equation}
(3r_{\rm i}-4r)\frac{1}{t_{\rm R}(r)}+2r(r_{\rm i}-r)\frac{d}{dr}\left(\frac{1}{t_{\rm R}(r)}\right)=3C\sqrt{r_{\rm i}-r},
\end{equation}
where $C$ is an arbitrary constant. In this equation, the first term on the left-hand side corresponds to the contribution reflected in Eq. \eqref{eq:domricc}, while the second term on the left-hand side corresponds to additional contributions to the Ricci scalar that show up when the radial dependence of the parameter $t_{\rm R}(r)$ is included. The right-hand side of this equation is the minimal expression which cancels the term that goes to zero in the denominator of Eq. \eqref{eq:domricc}, thus ensuring a good behavior of the Ricci scalar in the limit $r\rightarrow r_{\rm i}$. 

The inhomogeneous solution to this differential equation is given by the decreasing function
\begin{equation}
t_{\rm R}(r)=\frac{1}{C}\sqrt{r_{\rm i}-r}.\label{eq:dectr0}
\end{equation}
The simplest ansatz for our geometry is assuming that Eq. \eqref{eq:dectr0} holds for the entire interval $R_0\leq r\leq r_{\rm i}$. Demanding $t_{\rm R}(R_0)=t_{\rm P}$ fixes the constant $C$, leading to
\begin{equation}
t_{\rm R}(r)=t_{\rm P}\sqrt{\frac{r_{\rm i}-r}{r_{\rm i}-R_0}}.\label{eq:dectr}
\end{equation}
One can show by direct evaluation that this choice regularizes at the same time other curvature invariants such as, e.g., the Kretschmann scalar. Fixing $t_{\rm R}(r)$ as in Eq. \eqref{eq:dectr} specifies completely the geometry, in which the thin-layer region that encloses the non-standard ultraviolet effects that is depicted in Fig. \ref{Fig:transition} will be generally transformed into a (smoothed) triangular-shaped region, with one of the vertices located at $(t,r)=(0,r_{\rm i})$.

The inevitable introduction of the function $t_{\rm R}(r)$ is intimately related to the fact that non-perturbative ultraviolet effects are naturally confined in this case into a compact ball of radius $r_{\rm i}$, so that the form of the Schwarzschild metric is preserved for $r>r_{\rm i}$. In the most general case the maximum radius reached by ultraviolet effects could be an \emph{independent} parameter $r_{\rm m}$  such that $r_{\rm m}>r_{\rm s}$. The explicit construction of the corresponding geometries is more involved; we refer the reader to \cite{Barcelo2015} for a discussion of this issue. The geometries we have constructed explicitly have the necessary properties to reflect appropriately the main implications of the black to white hole transition in short characteristic time scales. We shall therefore take them as the starting point for our following discussions. 

\subsection{Short-lived trapping horizons \label{sec:5d}}

The first of the quantities of interest we can evaluate shows one of the nontrivial features that characterizes the entire family of bouncing geometries we are considering in this paper. In the first model we have discussed with a thin-layer transition region, calculations performed on the limit $t_{\rm R}\rightarrow 0$ will be exact up to $\mathscr{O}(t_{\rm R}/t_{\rm B})$ terms, thus implying this distributional limit will prove a very good analytical approximation on the basis of Eqs. \eqref{eq:treg} and \eqref{eq:tbounce}. This is also true in the second model with a triangular-shaped transition region, but now with $t_{\rm R}(R_0)\rightarrow 0$. Let us therefore start by considering the distributional limit $t_{\rm R}(R_0)\rightarrow 0$ for mathematical simplicity. In this case, the proper time for observers situated at $r=r_{\rm i}$ for the bounce to take place is exactly the same as for observers attached to the surface of the star. Indeed, from the perspective of an observer collapsing with the star, the entire process of collapse and bounce with respect to some reference initial position takes a time $t_{\rm B}$ as defined in (\ref{eq:tbounce}), that is, twice the free-fall collapsing time. By construction of the coordinate system, the proper time of the observer attached to the stellar structure is simply the Painlev\'e-Gullstrand time, $d\tau=dt$, as for that observer $dr/dt=v(t,r)$.  This same process seen by an observer always at rest at the initial position $r_{\rm i}$ takes {\em this very same time}. For this observer $dr=0$ and $v(t,r_{\rm i})=0$, implying $d\tau= dt$, so that its proper time is given the same temporal coordinate we are using to write down the metric. Seen by observers far outside the collapsing star ($r\gg r_{\rm s}$) the entire process would take this time multiplied by the standard general-relativistic redshift factor, $(1-r_{\rm s}/ r_{\rm i})^{-1/2}$. This redshift factor will be of order unity for stars initially larger than a few times their Schwarzschild radius. The fact that the proper time separation between these two events, the start of the collapse and its return to the initial position, is the same for the surface-attached observer and the one standing still at the initial radial position, is a general property of the geometries we are considering: a finite value of $t_{\rm R}(R_0)$ of the order of \eqref{eq:treg} leads, taking into account Eq. \eqref{eq:tbounce}, to extremely small $\mathscr{O}[t_{\rm R}(R_0)/t_{\rm B}]$ corrections coming from the finite transient zones. Therefore, these geometries do not exhibit long-lived trapped regions of any sort, but only short-lived trapped regions.

From a different point of view, imagine that one were to monitor with high time resolution this time-symmetric collapse from the reference initial position, that is assumed to be sufficiently far outside the Schwarzschild radius. One sets up two synchronized clocks at this initial position before the collapse. Then one clock is left to follow the collapsing structure and the other is kept at rest in the reference position. By observing with a telescope the ticks of the clock falling with the star, one would see that in the collapsing phase the ticks slow down progressively. However, at some point they start to speed up in such a way that when the two clocks are finally back together they show precisely the same time. This is easily understood if we think in terms of the analogue metric in fluids \cite{Barcelo2005}. When the star is collapsing, the velocity profile is that of a sink, so that signals originated on the star's surface will be emitted at different positions, picking up a delay that depends on the time that light needs to cover up that distance. On the contrary, in the white-hole patch the velocity profile is reminiscent of a source, which effectively shortens the time between different signals for the external observer. Overall, for an external observer, that uses essentially the Schwarzschild time coordinate as its proper time (but for an irrelevant redshift factor depending on his position), the collapsing phase will last longer than the expanding phase, so that he might not realize the time-symmetric character of the process.

In summary, one of the essential features of the process we are considering is that the time lapse associated with the collapse is short, of the order of tenths of a millisecond for neutron-star-like initial configurations. This is equally true both for observers attached to the structure as well as for external stationary observers. Within this quite general family of geometries (arguably a complete set of geometries interpolating between a black-hole geometry with the white-hole geometry), the only way to prescribe geometries that allow for extremely long times of the bounce process as seen by external observers is to introduce extremely slow transients. By using the adjective \emph{extremely slow} we understand a regularization that takes a very long time, measured by observers attached to the surface of the star, in order to overcome the gravitational attraction and start a noticeable expansion  of the stellar structure; in mathematical terms this corresponds to very large $t_{\rm R}(R_0)/t_{\rm B}$ quotients. From the perspective of the regularization of the singularity as well as observers \emph{inside} the stellar structure, this quasi-static possibility is arguably quite unnatural. In the framework of our discussion in Sec. \ref{sec:4}, these solutions would correspond to prioritizing the role of the external metric in the entire process, subjugating the behavior of matter as well as the possible ultraviolet effects to the prevalence of the long-lived horizon. 

\subsection{Short transients and the propagation of non-perturbative ultraviolet effects \label{sec:5f}}

All the transients lead to a characteristic imprint as we have already discussed: the deviation of the near-horizon outer geometry, that is, the Schwarzschild geometry beyond $r=r_{\rm s}$. Such a feature is anathema in the orthodox view on the possible relevance of ultraviolet effects on classical geometries. It belongs to the conventional wisdom that appreciable deviations from the classical behavior are to be expected in regions of large curvature in, e.g., Planck units. The argument is that only then the possible corrections to the Einstein field equations are expected to be non-perturbative. In the Schwarzschild solution the Ricci curvature tensor is zero but the Weyl part of the Riemann curvature tensor leads to the Kretschmann scalar
\begin{equation}
\mathscr{K}=\mathscr{R}_{abcd}\mathscr{R}^{abcd}=\frac{12 r_{\rm s}^2}{r^6}.\label{eq:kre}
\end{equation}
This implies that the geometry around $r=r_{\rm s}$ of stellar-size black holes is to be regarded as a robust feature, not to be affected by any kind of ultraviolet modifications.

While this view certainly makes sense in the study of the static case (i.e., an eternal black hole), the role of these considerations is far from clear for us when talking about dynamical situations. For instance, in our proposal there is no modification at all of the behavior of the geometry near the moment of formation of the trapping horizon in the black-hole patch. An observer who suddenly left the surface of the star to pend close, but outside the horizon, will not notice any deviations from the expected behavior in GR in the few initial instants of the process. It is only when matter reaches a dense enough situation that relevant ultraviolet effects altering the geometry are originated. While at first these effects are confined to regions of high curvature, there is nothing that prevents them to propagate outwards, modifying the geometry in their wake. In doing so these effects can reach regions that were characterized by low curvatures, such as the near-horizon region. However, this effect should not be seen as a modification of the behavior of regions with low curvature by unnaturally large ultraviolet effects, but rather as the propagation of a non-perturbative wave of high-curvature through regions of low curvature. Interestingly, this picture is self-consistent only for short transients, with characteristic time scale $t_{\rm R}(R_0)$ of the order of the Planck time.

This is neatly illustrated by taking as working example the geometry considered in Sec. \ref{sec:5a}, though the situation is generically the same for all the geometries we have considered. The Kretschmann scalar of the transient in this case is given in \eqref{eq:kremod}. As argued above, the short (Planckian) time $t_{\rm R}$ associated with the transient implies that the modifications of the spacetime curvature in the surroundings of the stellar structure, when the latter reaches its minimum radius $R_0$, are also of the order of the Planck curvature. Thus in geometries with short transients, and only in these cases, the trigger of non-perturbative ultraviolet effects on the metric outside the star is inextricably tied up to spacetime regions with high curvature. This region of high curvature will propagate outwards, getting diluted in this process: already when reaching the horizon, the magnitude of this curvature several orders of magnitude lower, roughly by
\begin{equation}
\left(\frac{R_0}{r_{\rm s}}\right)^{3/2}\sim\frac{m_{\rm P}}{M}.
\end{equation}
One might be tempted to argue on this basis that ultraviolet effects should not be able to significantly alter the near-horizon geometry, being the corresponding curvature far smaller than Planckian.\footnote{It is interesting to note that it is in principle possible to construct models in which the consideration of triangular-shaped transition regions could even lead to Planckian curvatures near the horizon, by just making an appropriate ansatz for $t_{\rm R}(r)$ from $r=R_0$ up to the near-horizon region.} The situation is, indeed, the contrary. While the classical distribution of the curvature (measured by the Kretschmann scalar) decays with the radius as $r^{-3}$ [Eq. \eqref{eq:kre1}], the curvature of the non-standard region associated with ultraviolet effects decays as $r^{-3/2}$ [Eq. \eqref{eq:kremod}]. This guarantees that modifications that are non-perturbative at $R_0$, remain non-perturbative near the horizon. Thus modifications of the near-horizon geometry do not appear because of unnaturally large ultraviolet effects originated there, but rather as a result of the propagation of sudden ultraviolet effects that are originated when the stellar structure undergoes a violent bounce at Planckian densities.

It is certainly not possible, given our present knowledge of the gravitational interaction, to back up the picture with short characteristic time scales from a first-principle, complete perspective. The work of {H{\'a}j{\'{\i}}cek} and collaborators \cite{Hajicek2001,Hajicek2003,Ambrus2004,Ambrus2005}, represents nevertheless a first step in this direction. In their study of the quantization of collapsing shells, they have shown that the transition from a collapsing branch to an expanding branch occurs, and that a given definition of the crossing time between an external observer and the dynamical shell is short, being essentially of the same order as the time $t_{\rm B}$ we are considering, given by Eq. \eqref{eq:tbounce}. It is equally worth remarking their skepticism about this result, as the same authors were expecting very long times to occur, in order for their model to be accommodated within the REBH family. The definition of observables concerning time intervals in these bouncing processes is however far from straightforward \cite{Ambrus2004,Ambrus2005}. At the light of the developments presented here, this issue certainly deserves further study, both in this particular model and in frameworks that use different quantization techniques, such as \cite{Hartle2015} for instance. 

In any case, lacking for the moment a fundamental justification for the short transients does not prevent our discussion to be self-consistent at low energies, while presenting new intuitions about the way in which ultraviolet modifications to the behavior of GR would behave. On top of this, what we want to stress is the genuine opportunity that these hypothetical processes offer: in contrast with REBH proposals, which are experimentally inert for extremely long times and thus for practical purposes, a process with a short characteristic time scale should lead to clear imprints that could be hopefully detectable. Thus this proposal is audacious, but not without consequences, as it offers prospects of being falsifiable much more easily than any other models that are nowadays present the literature.

\section{Physical and observational consequences \label{sec:6}}

If there exists a regularization of the classical behavior of the form we have described, the collapse process itself would not constitute the final stage of collapse in stellar physics. One would immediately be impelled to wonder about what would happen after the bounce. The search for new states of equilibrium on the one hand, and the understanding of the transient collapse process itself on the other, become entirely distinct issues. 

If our idea is at work in nature it should have many observable effects, though additional work is needed in order to determine the possible signatures of both phases. In the collapse of, say, a neutron star, matter would remain apparently frozen at the Schwarzschild radius for just a few tenths of a millisecond before being expelled again. Even neglecting dissipation, the metric is non-Schwarzschild during an extremely short time interval in a region extending outside the gravitational radius. In realistic situations the bounce will not be completely time-symmetric: part of the matter will go through towards infinity in the form of dissipative winds while the remaining mass will tend to recollapse. In this way one would have a brief and violent transient phase, composed of several bounces, followed by the formation of a new (stable or metastable) equilibrium object with the resulting mass.

\subsection{Towards new figures of equilibrium \label{sec:6a}}

In an ideal situation, perfectly spherically symmetric and without dissipation, the collapsing body would enter into a never-ending cycle of contracting and expanding phases. In a realistic situation, though, one expects that the system will dissipate at least quantum mechanically while searching for new equilibrium configurations. Here the panorama of possibilities is almost unexplored. Let us discuss the possibility we think more plausible. Notice that any development on this direction should properly acknowledge and deal with, for instance, the observations of the existence of extremely dense objects in our own galaxy.

In~\cite{Barcelo2008} it was shown that if the velocity of trapping-horizon crossing by the collapsing matter distribution were rather small, then quantum effects of vacuum polarization would become so powerful that they might even stop the collapse. It is very unlikely that the classically expected almost-free-falling collapse of stellar structures like neutron stars would lead directly to strong vacuum polarization effects. However, in our scenario, when taking into account dissipation, one would expect that each new recollapsing phase would start from a position closer to its Schwarzschild radius than the previous one. In this way, at some stage vacuum polarization effects could start to be relevant and even stop further collapses. This might lead for the time being to hypothetical almost-stationary structures hovering extremely close to their gravitational radius. 

There might be other mechanisms underlying the stabilization of stellar structures close to their gravitational radius. What is relevant here is that the final metastable object could be small, dark, and heavy, but without black- or white-hole districts (see~\cite{Visser2008} and references therein). These black stars will not be voids in space, but they will be filled with matter. Since they have no horizons, they will in principle be completely open to astrophysical exploration. Let us stress that black holes as described by classical GR might still continue to be very good models for the external gravitational behavior of these black stars.

A natural question in this regard is whether Hawking-like evaporation, being a paradigmatic theoretical feature of black holes, would be also a characteristic of these new objects. When the system stabilizes close to its Schwarzschild radius, it might emit or not and with different spectral properties depending on the specifics of the structure, which at this stage are difficult to envisage. In other words, the Hawking effect \emph{does not} need to be preserved in the black-star scenario. However, we have already mentioned in our previous discussion that there are ways in which these objects could acquire emission and evaporation properties resembling Hawking's scenario. At least two different mechanisms are known. One such structure could emit a Hawking-like flux if it were continuously and asymptotically approaching its Schwarzschild radius (without crossing it)~\cite{Barcelo2006}, or if it were pulsating in a close to free-fall manner~\cite{Barbado2011}. Should this radiation exist, it would in both cases be Planckian but not strictly thermal, as correlations are maintained by both mechanisms, and the total amount of energy radiated would be in principle negligible given the extremely short time associated with the transients. In relation with Hawking radiation, notice that these scenarios do not invite us to wonder whether information is lost or not, as no singularities and no long-lived trapping horizons are formed in the first place. On the other hand, during the transient phases one would expect quantum dissipation in the form of particle production. This particle production will be in general non-thermal, though at trapping-horizon crossings it would have the form of short bursts of thermal radiation~\cite{Stephens1994,Barcelo2006,Barbado2011}. 

Once the object has settled down, it would probably be extremely difficult to distinguish it from a standard black hole through astrophysical observations. There have been some proposals to discern whether or not there exists an explorable surface in objects associated with black holes \cite{Narayan2002,Narayan2008}. The absence of Type I X-ray emission in binaries containing a black-hole mimicker have been argued to imply that the candidate did not have an external surface but an event horizon~\cite{Narayan2002}. However, things are clearly not that simple (for some specific criticism see~\cite{Abramowicz2002}). The reaction of the black hole mimicker when absorbing some matter from its companion would strongly depend on its specific heat capacity. If the black-hole candidate has a heat capacity similar to those of a black hole, which can be expected due the high redshift of its surface, its behavior would be difficult to distinguish in this regard from that of a proper black hole. This kind of observations can play a significant role, though, in putting constraints to specific models of black hole mimickers (see, e.g.,~\cite{BroderickNarayan2007} for constraints on gravastars). This task could even be more difficult for proposals in which the compact object lacks a hard surface \cite{Vincent2015}.

A different issue is the possibility, in principle, of the hypothetical use of a radar to check the presence or not of a surface. As discussed in~\cite{Barcelo2015}, a \emph{elastic} scattering of a wave in the supposed surface would distinguish in no time whether a surface exists or not.  

\subsection{Energetics of the transient phase \label{sec:6b}}

When considering realistic situations with dissipation, the transient phase might leave some traces, for instance in the physics of gamma-ray bursts (see, e.g., \cite{Piran2004}). There is experimental evidence that a subset of these events, the so-called \emph{long} gamma-ray bursts, are associated with the final stages in the life of very massive stars. The most widely accepted theoretical picture is known as the collapsar model \cite{MacFadyen1998}. It is natural to expect that a modification of the standard gravitational collapse process to a black hole that is considered here could leave clear imprints associated with a reverberant collapse. However, in the collapsar model of GRBs the emission zone is supposed to be very far from the collapsed core \cite{Piran2004}. This means that the connection between the processes at the core and those at the external wind shells could be very far from direct. However, the general features of the model are enough in order to roughly compare its energetics to those of GRBs. This comparison may be used in order to understand whether or not the bounce process is a reasonable candidate for the mechanism behind these bursts. So let us assume that the picture discussed above is realized in nature: the occurrence of violent bouncing processes, dissipation and final stabilization in the form of a black star. We can estimate the effect of the energy loss in the entire process by using the following argument. Recall that for dust matter content, and in the absence of rotation, the differential equation for the trajectory of the surface of the star is mathematically equivalent to that of a test particle with the overall mass of the star following a radial geodesic in Schwarzschild spacetime. We can then use the conserved quantities associated with the geodesic equations in this spacetime. In particular, we shall use the conserved quantity $E$ that is associated with energy. So let $r_{\rm i}$ be the initial radius and $r_{\rm s}$ the Schwarzschild radius of the star, and consider the Schwarzschild effective potential for radial motion. If the structure was originally at rest, its energy is given by
\begin{equation}
\left(\frac{E}{M c^2}\right)^2=1-\frac{r_{\rm s}}{r_{\rm i}}.\label{eq:schen}
\end{equation}
This equation has a clear interpretation: the positive term on the right-hand side is the rest energy of the star, while the second term corresponds to the negative gravitational energy of the structure. Then the energy of the resulting compact body, as defined in \eqref{eq:schen}, is essentially zero. Energy balance implies that the energy that has to be released in the entire process, for example by means of the emission of a shell of matter that escapes to spatial infinity, is given by:
\begin{equation}
\Delta E=Mc^2\sqrt{1-\frac{r_{\rm s}}{r_{\rm i}}}.\label{eq:energetics}
\end{equation}
This is a model-independent estimation that tell us that the object has to get rid of a significant portion of its original rest energy in order to reach stabilization. If we take for example a neutron star with $r_{\rm i}=2\,r_{\rm s}$, then the emitted energy is $Mc^2\sqrt{1/2}\simeq 0.71\, Mc^2$.

How is this compared with the GRBs energetics? Interestingly, \eqref{eq:energetics} is of the same order of magnitude as the energy emission in those events if the emission is considered isotropic (see, e.g., \cite{Piran2004,Mobberley2009}). Indeed, it is of the same order of magnitude as other theoretical estimates that are based, for instance, on the Penrose process \cite{Frolov2011} or similar mechanisms for energy extraction from charged black holes \cite{Christodoulou1971,Damour1975}. A comment that applies to all these estimations is that there is experimental evidence that the emission is collimated so that the real energy that is emitted is smaller than \eqref{eq:energetics}, roughly by a factor of $10^{-2}$ . Of course, \eqref{eq:energetics} is a crude estimate that does not take into account other effects that would take place near the collapsing object, besides being evaluated in an isotropic model that does not take into account rotation. That with simple, model-independent ingredients we are able to get that close to the observed energetics of GRBs is a strong incentive to consider further developments of the picture. This is a hint in favor of the possibility that the bounce process we are describing might be behind of some GRB events. As we cover in the next section, gravitational waves produced deep inside these violent events may provide a much better observational opportunity, eventually permitting to elucidate the kind of mechanism that is behind them.

\subsection{Ripples from the transient phase \label{sec:6c}}

One of the primary predictions of GR that still awaits experimental corroboration is the generation and propagation of local disturbances of spacetime, or gravitational waves. Gravitational-wave astronomy is nowadays a mature branch from a theoretical perspective, while there has been a great deal of experimental effort in order to overcome the difficulties in the detection of these tiny ripples in spacetime. The scientific potential that this new observational window promises is huge; see \cite{Fryer2011,Buonanno2014} for instance. The fine theoretical knowledge of the gravitational wave patterns associated with different gravitational phenomena would make possible to unveil information about astrophysical processes that is definitely not possible to obtain from their electromagnetic counterparts. As a prime example, this observational technique is arguably the best tool to finally determine the physical mechanisms that are behind both short and long GRBs. If the process that we are describing in this paper is realized in nature, the information encoded in gravitational waves may be significantly different, and therefore more surprising than initially expected.

The observation of the gravitational wave pattern associated with the gravitational collapse of a massive star into a black hole, if properly correlated with the electromagnetic counterpart of a long GRB, would be the smoking gun of the collapsar model. The generation and form of these wave patterns are well understood nowadays. In the spherically symmetric case, non-spherical inhomogeneities that are present in the initial stellar structure will generate a gravitational wave signal, that terminates with the relaxation of the perturbed horizon to its stable Schwarzschild form (in the presence of rotation it will be described by the Kerr solution instead \cite{Hawking1973}). After that, there is complete silence in the gravitational-wave channel. This is a definite characteristic of the classical gravitational collapse process as described in GR, that will be shared by \emph{all} the REBH models. The robustness and generality of this result is what makes any departures from it highly interesting. We shall describe in the following why do we expect distinctive departures of this behavior to occur in our model, and how can these be computed. It is noteworthy that this is only proposal in the literature to our knowledge that advocates for these kind of modifications on the gravitational wave patterns of collapsing stars.

Let us start with a very simple analogue, given by the electromagnetic radiation of a pulsating, or bouncing, charged sphere. In the spherically symmetric case there will no emission of radiation. As in the gravitational case, one has to consider non-spherical distributions of charge in order to trigger the emission of electromagnetic radiation. Now the electromagnetic radiation measured far from the sphere will depend on the given trajectory of the radius of the sphere, $R(t)$, bounded both from below and above. We take this trajectory as the analogue of the trajectory of the star surface in our bouncing geometry. The overall emission of radiation can be obtained in the framework of standard electrodynamics. The emission of electromagnetic radiation in the collapsing branch will be followed by the emission of new radiation pulses coming from the expanding branch, perhaps with a burst corresponding to the bounce event. Notice that the emission of radiation breaks the time symmetry of the trajectory $R(t)$, if present. Additionally, in several bounces we shall get several repetitions of this wave pattern. This is what we expect to occur in the gravitational case, thus leading to a very different gravitational signal when compared with the one obtained in GR.

Now when we turn back to gravity, there is a crucial technical difference to take into account when considering this analogy. While the equations of electrodynamics enforce the conservation of charge, the equations of GR do the equivalent with energy. In the electromagnetic case the energy that is emitted in the form of radiation is ultimately provided by the mechanism behind the kinematic evolution of the charge distribution. In other words, the non-spherical distributions of charge that cause the emission of radiation do not decay over time. On the contrary, the equations of GR automatically take into account the decay of non-spherical perturbations due to the emission of gravitational waves. This makes the problem more involved mathematically but, as we discuss in the following, having at hand the metric describing the bounce  of the stellar structure one should be able to obtain a definite answer for the spectrum of gravitational waves that is produced in the process for a given perturbation of the initial configuration.

Let us sketch the necessary steps in order to do so, taking for instance the metric described in Sec. \ref{sec:5c} as a specific representative of the bounce process. This metric is a solution of the field equations
\begin{equation}
G_{ab}=\frac{8\pi G}{c^4}\left(T_{ab}+\mathscr{T}_{ab}\right),\label{eq:eins}
\end{equation}
where $G_{ab}$ is the Einstein tensor, $T_{ab}$ is the dust stress-energy tensor, and $\mathscr{T}_{ab}$ is the non-standard stress-energy tensor, the components of which would be similar to \eqref{eq:effset}, that describes the shock wave produced in the violent bounce. Our previous discussion corresponds to the spherically symmetric situation; in order to produce gravitational waves we need to introduce non-spherical perturbations. These perturbations are introduced in the initial state of the system, namely the star at rest with radius $r_{\rm i}$, with asymptotic conditions ensuring the absence of radiation at spatial infinity. We can perform a perturbative expansion around the spherically symmetric situation of the form
\begin{equation}
G_{ab}=G_{ab}^{(0)}+G_{ab}^{(1)}+...,
\end{equation}
where $G_{ab}^{(0)}$ is the Einstein tensor of the spherically symmetric geometry describing the bounce, and $G_{ab}^{(1)}$ will contain the information about the non-spherical perturbations and, in particular, the gravitational wave signal. At first order in the deviations from absolute spherical symmetry, the evolution of these perturbations are then determined by the equations
\begin{equation}
G^{(1)}_{ab}=\frac{8\pi G}{c^4}T_{ab}^{(1)}.\label{eq:einspert}
\end{equation}
Notice that $\mathscr{T}_{ab}$ drops off from this equation, as it is constructed to identically cancel the zeroth-order terms. This last equation sets the basis for the study of gravitational wave emission of this process. The study of its implications is currently being carried out, the results of which will be reported elsewhere.

It is probably not necessary to stress the appeal of the possibility, even if it could appear remote at present, of detecting a characteristic gravitational wave signal that deviates from the expected classical template. Such an observation would be the smoking gun of the bounce process as described here, providing a low-energy observational window to genuine ultraviolet effects acting on the gravitational collapse of massive stars.

\section{Conclusions \label{sec:7}}

The outcome of extreme gravitational collapse processes is one of the great theoretical open problems in gravitational physics. The precise determination of the ultraviolet modifications to the classical behavior encapsulated in GR is of course the essential key to unveil its solution. It is not so often stressed that the present understanding of this problem is facing an important dilemma: most of the models in the market largely preserve the semiclassical picture of long-lived trapping horizons, thus obstructing their very experimental verification due to the ridiculous large lifetime of any REBH model. While we are far from denying the theoretical value of these models in understanding the gravitational interaction, in our opinion a serious effort should be made in analyzing the theoretical and observational characteristics of models that exhibit clear testable signatures in the near future. In this paper we have reviewed and worked out new properties of one such alternative model to REBHs.

While the perturbative view of the classical picture that is encoded in the REBH paradigm certainly represents the consensus of the community, we have tried to transmit that there exists an interesting alternative. In this alternative model, ultraviolet effects are no longer perturbative, so that the regularization of the singular behavior of GR opens new unexplored avenues for the evolution of the system. The resulting geometries cannot be described as representing essentially the semiclassical perturbation of the trapping horizons that are formed in the collapse, but give preeminence to the regularization of the singularity. In plain terms, these geometries represent the bounce of the matter distribution when reaching Planckian densities, originating a shock wave that propagates outwards, modifying the near-horizon Schwarzschild geometry. All these features have been neatly justified through the careful construction of the explicit form of the corresponding geometries. One of the new general conclusions reached in this paper, by means of the study of curvature invariants of these geometries, is that only for extremely short transients (of the order of the Planck time) the non-perturbative modification of the near-horizon Schwarzschild geometry is justified by the propagation of non-perturbative ultraviolet effects originated at the moment of the bounce.

Given the singular character of this model, it provides a unique opportunity for further studies of unexplored issues. For instance, the suggestions it presents about the possible low-energy effects coming from the ultraviolet gravitational regime, and the characteristic time scales for the black to white hole transition, are worth contemplating from a first-principles perspective. On the other hand this model has the potential, if getting mature enough, to lead to surprises for our conception of astrophysical black holes. As we have tried to remark, the most relevant property that singles out this proposal within the vast literature in the subject is its overall short characteristic time scale, which makes possible to think about the (always exciting) possibility of making contact with the empirical reality.


\acknowledgments
Financial support was provided by the Spanish MICINN through the projects FIS2011-30145-C03-01 and FIS2011-30145-C03-02 (with FEDER contribution), and by the Junta de Andaluc\'{\i}a through the project FQM219. R.C-R. acknowledges support from CSIC through the JAE-predoc program, co-funded by FSE.

\bibliographystyle{unsrt}
\bibliography{references}

\end{document}